\documentstyle[preprint,aps]{revtex}
\begin{document}
\title{Formulation of the Classical Mechanics in the Ring of Operators}
\author{A. Ver\c{c}in}
\address{Department of Physics \thanks{%
Permanent address}\\
Ankara University, Faculty of Sciences,\\
06100, Tando\u gan-Ankara, Turkey\\
E.mail:vercin@science.ankara.edu.tr\\
and\\
Feza G\"{u}rsey Institute\\
P.O.Box 6, 81220 \c{C}engelk\"{o}y\\
\.{I}stanbul, Turkey\\
E.mail:vercin@fge1.gursey.gov.tr \\}
\maketitle

\begin{abstract}
By making use of the Weyl-Wigner-Groenewold-Moyal association rules, a
commutative product and a new quantum bracket are constructed in the ring of
operators ${\cal {F}(H)}$. In this way, an isomorphism between Lie algebra of
classical observables (with Poisson bracket) and the Lie algebra of quantum
observables with this new bracket is established. By these observations, a
formulation of the classical mechanics in ${\cal {F}(H)}$ is obtained and is
shown to be $\hbar\rightarrow 0$ limit of the Heisenberg picture formulation
of the quantum mechanics.
\end{abstract}

\tightenlines

\section{Introduction}

In this report we are going to answer, in a most general setting, two
related questions: (1) What is the analogue of the multiplicative structure
of the classical observables (functions defined on a classical phase space)
in the quantum formalism?, (2)``What is the image of the Poisson bracket
(PB) of functions in the ring of operators''(of quantum observables)? These
questions, in that way or another, were in the mind of many physicists since
the very beginning days of quantum mechanics \cite{Dirac,Gotay} and, to the
best of my knowledge, they are not answered yet. Especially the second
question was explicitly stated, as is quoted here, in the second of two
seminal papers of Ref. \cite{Bayen} and in a figure of which an empty box
was used for the image of the PB. Throughout this paper we assume that the
phase space is ${\bf R}^{2d},d$ integer.

In the well-known canonical quantization, for the analogue of the
multiplication ``product $\rightarrow $ anti-commutator'' rule works well up
to cubic polynomials. For the analogue of PB `` PB $\rightarrow (i\hbar)^{-1}
$ commutator ([,])'' rule works well up to quadratic polynomials of position
and momentum variables, and up to observables which are affine functions of
the position or of the momentum. According to the Groenewold-Van Howe
theorem they lead to inconsistencies for the quartic and cubic polynomials,
respectively. Similar obstructions arise for some other phase spaces which
have different topology from ${\bf R}^{2d}$. For more extensive and
technical discussion of this topic we refer to Ref. \cite{Gotay,Folland} and
references therein.

These two questions, which underline the fundamental differences between the
classical mechanics and quantum mechanics, will be answered by making use of
the Weyl-Wigner-Groenewold-Moyal (WWGM)-quantization scheme (for recent
reviews see \cite{Hillery,Sternheimer}). The WWGM-quantization enables us to
carry the quantum theory to a phase space and giving it an autonomous
structure \cite{Bayen} with its own ``genvalue" equations \cite{Fairly},
(quasi)probability distributions \cite{Hillery} and spectral resolution.
Quantum information encoded in the noncommutative product of the quantum
observables is transferred via the WWGM-association to classical phase space
and stored in the noncommutative $\star-$product (see Eq. (16) below) of the
classical observables. In this way, to the product of operators corresponds
the $\star-$product of functions and to the commutator of operators
corresponds the Moyal bracket (MB) of functions. The resulting theory is
also referred as the deformation quantization, or as the phase space
formulation of the quantum mechanics. On the other hand, this quantization
scheme also enables us to carry the classical mechanics on a phase space to
a Hilbert space. This paper concentrates on this latter aspect of this
quantization, although almost all the literature deals with the former. More
concretely, we search for what correspond to the commutative product of
functions and to the PB of them in the WWGM-quantization. Our answers to
these questions will lead us to a formulation of the classical mechanics in
the ring of operators.

For the purposes of this report we mainly consider systems with one degree
of freedom ($d=1$) and the corresponding phase spaces in real coordinates.
Generalizing our results to systems with finite or denumerably infinite
number of degrees of freedom and to phase spaces with complex coordinates
are straightforward. Such a generalization of one of the main results of
this paper will be given at the end of section IV. We use the
derivative-based approach developed in \cite{Vercin}, which is different
from the conventional integral-based one but can be considered as a
Liouville space formalism \cite{Royer}. In this formalism operators are
represented by (super)kets and superoperators (see Section II below) act on
them. The derivative based approach shows that when operators are labeled by
some parameters; derivatives with respect to them, multiplication of the
operators with them and even integration over them must be considered as
superoperators. This is crucial point of the WWGM-quantization for it comes
into play by considering the parameters of the group space as the coordinate
functions of a phase space, and these parameters are carried by the used
operator basis as labels.

The organization of the paper is as follows. In Section II, we define some
important superoperators which are commutative function of their arguments.
The importance of these superoperators is made manifest in Section III,
which includes a brief review of some fundamental ideas of the
WWGM-quantization, and our answer for the first question. As the second main
result of this paper, a new quantum bracket is derived in Section IV. In
Section V we give some general applications by using this new bracket. We
conclude with a brief summary and discussion of results.

\section{Liouvillian Superoperators}

Let us consider the Heisenberg-Weyl (HW) algebra: $[\hat q,\hat p]=i\hbar%
\hat I$, where $\hbar =h/2\pi ,\hat I,\hat q$ and $\hat p$ are the Planck's
constant, the identity operator and the Hermitian position and momentum
operators, respectively. Here and henceforth operators and functions of
operators acting in ${\cal {H}}$ are denoted by $\hat{}$ over letters and
superoperators by $\hat{}$ over boldface letters. In terms of HW-algebra and
a complex parameter $s\in {\bf C}$ we define the superoperator 
\begin{eqnarray}
\hat{\bf O}^{(s)}_{nm}=2^{-(n+m)}\hat{\bf T}^{n}_{[\hat{q}]_{(s)}}
\hat{\bf T}^{m}_{[\hat{p}]_{(-s)}}
\end{eqnarray}
where, $\hat {{\bf L}}_{\hat A}$ and $\hat {{\bf R}}_{\hat A}$ being,
respectively, multiplication from left and from right by $\hat A$ 
\begin{eqnarray}
\hat{\bf T}_{[\hat{A}]_{(s)}}=(1+s)\hat{\bf L}_{\hat{A}}+
(1-s)\hat{\bf R}_{\hat{A}}.
\end{eqnarray}
Note that for an arbitrary operator $\hat F$ 
\begin{eqnarray}
[\hat{\bf T}_{[\hat{q}]_{(s)}} ,\hat{\bf T}_{[\hat{p}]_{(-s)}}]\hat{F}=0.
\end{eqnarray}
The actions of $\hat {{\bf O}}_{nm}^{(s)}$ on the unit operator $\hat{I}$
and, under the trace sign, on an arbitrary operator $\hat F$ are as follows; 
\begin{eqnarray}
\hat{\bf O}^{(s)}_{nm}(\hat{I})&=&\hat{t}^{(s)}_{nm}, \\
Tr[\hat{\bf O}^{(s)}_{nm}(\hat{F})]&=&Tr[\hat{t}^{(-s)}_{nm}\hat{F}],
\end{eqnarray}
where 
\begin{eqnarray}
\hat{t}^{(s)}_{nm}&=&2^{-(n+m)}\hat{\bf T}^{n}_{[\hat{q}]_{(s)}}
\hat{\bf T}^{m}_{[\hat{p}]_{(-s)}}\hat{I} \nonumber\\
&=&2^{-(n+m)}\hat{\bf T}^{m}_{[\hat{p}]_{(-s)}}
\hat{\bf T}^{n}_{[\hat{q}]_{(s)}}\hat{I}, \\
&=&2^{-n}\sum^{n}_{j=0}(^{n}_{j})(1+s)^{j}
(1-s)^{n-j}\hat{q}^{j}\hat{p}^{m}\hat{q}^{n-j} \nonumber\\
&=& 2^{-m}\sum^{m}_{k=0}(^{m}_{k})(1-s)^{k}
(1+s)^{m-k}\hat{p}^{k}\hat{q}^{n}\hat{p}^{m-k},
\end{eqnarray}
is the $s$- ordered product of a term containing $n$ factors of $\hat q$ and 
$m$ factors of $\hat p$. In obtaining these expressions we note that
ordering parameters $-s$ and $s$ in the last factors of the first two lines
of the above expressions do not contribute to the results since $\hat {{\bf T%
}}_{[\hat A]_{(\pm s)}}^m\hat I=2^m\hat A^m$. Moreover, in the last two
lines we made use of the binomial formula 
\begin{eqnarray}
\hat{\bf T}^{n}_{[\hat{A}]_{(s)}}
=\sum^{n}_{j=0}(^{n}_{j})(1+s)^{j}(1-s)^{n-j}
\hat{\bf L}^{j}_{\hat{A}}\hat{\bf R}^{n-j}_{\hat{A}}.
\end{eqnarray}

From Eqs. (6) and (7) we have, for $s=\pm 1$, 
\begin{eqnarray}
\hat{t}^{(1)}_{nm}=\hat{\bf L}^{n}_{\hat{q}}\hat{\bf R}^{m}_{\hat{p}}\hat{I}=
\hat{q}^{n}\hat{p}^{m}, \qquad \hat{t}^{(-1)}_{nm}=
\hat{\bf L}^{m}_{\hat{p}}\hat{\bf R}^{n}_{\hat{q}}\hat{I}=\hat{p}^{m}\hat{q}^{n},
\end{eqnarray}
and for $s=0$, 
\begin{eqnarray}
\hat{t}^{(0)}_{nm}=2^{-n}\sum^{n}_{j=0}(^{n}_{j})\hat{q}^{j}\hat{p}^{m}
\hat{q}^{n-j}
=2^{-m}\sum^{m}_{k=0}(^{m}_{k})\hat{p}^{k}\hat{q}^{n}\hat{p}^{m-k}.
\end{eqnarray}
While relations (9) exhibit the standard ($s=1$) and antistandard ($s=-1$)
rule of ordering, that corresponding to $s=0$ are two well known expressions
of the Weyl, or symmetrically ordered products. In fact, the usual
expression known for the Weyl ordered form of $\hat t_{nm}^{(0)}$ is a
totally symmetrized form containing $n$ factors of $\hat q$ and $m$ factors
of $\hat p$, normalized by dividing by the number of terms in the
symmetrized expression. Here we give an example 
\begin{eqnarray}
\hat{t}^{(0)}_{12}=\frac{1}{2}(\hat{q}\hat{p}^{2}+\hat{p}^{2}\hat{q})
=\frac{1}{4}(\hat{q}\hat{p}^{2}+2\hat{p}\hat{q}\hat{p}+\hat{p}^{2}\hat{q})
=\frac{1}{3}(\hat{q}\hat{p}^{2}+\hat{p}\hat{q}\hat{p}+\hat{p}^{2}\hat{q}).\nonumber
\end{eqnarray}
As a simple result of the approach followed here, explicit expressions for
many forms of the $s-$ordered products and their equivalences, without using
the usual commutation relations, naturally arise by noting only the relation
(3). From (7) easily follows that $[\hat t_{nm}^{(s)}]^{\dagger }=\hat t%
_{nm}^{(-\bar s)}$, that is, for general $n,m$ integers, $\hat t_{nm}^{(s)}$
are Hermitian if and only if $\bar s=-s$ ($\bar s$ denotes the complex
conjugation of $s$). In particular, the Weyl ordered products $\hat t%
_{nm}^{(0)}$ are Hermitian. Since the result is independent of $s$ when both
or one of the integers $n$, $m$ is zero, these special monomials are
Hermitian for any value of $s$.

Because of Eqs. (4) and (5) we would like to call $\hat{{\bf O}}^{(s)}_{nm}$
the ordering superoperator. Making use of (1) and (5) we obtain the following
relations for their repeated actions 
\begin{eqnarray}
\hat{t}_{nm}^{(s)}&=&\hat{\bf O}^{(s)}_{nm}(\hat{I})=
\hat{\bf O}^{(s)}_{n_{1}m_{1}}(\hat{\bf O}^{(s)}_{n_{2}m_{2}}(\hat{I}))=
\hat{\bf O}^{(s)}_{n_{1}m_{1}}(\hat{t}^{(s)}_{n_{2}m_{2}}), \\
Tr[\hat{\bf O}^{(s)}_{nm}(\hat{F})]&=&Tr[\hat{\bf O}^{(s)}_{n_{1}m_{1}}
(\hat{\bf O}^{(s)}_{n_{2}m_{2}}(\hat{F}))]\nonumber\\
&=&Tr[\hat{t}^{(-s)}_{n_{1}m_{1}}
(\hat{\bf O}^{(s)}_{n_{2}m_{2}}(\hat{F}))]\nonumber\\
&=&Tr[\hat{F} \hat{\bf O}^{(-s)}_{n_{2}m_{2}}
(\hat{t}^{(-s)}_{n_{1}m_{1}})],
\end{eqnarray}
where $n=n_{1}+n_{2}, m=m_{1}+m_{2}$.
Here is an example of the relation (11)
\begin{eqnarray}
\hat{t}^{(s)}_{n+1, m+1} &=& \frac{1}{4}\hat{\bf T}_{[\hat{q}]_{(s)}}
\hat{\bf T}_{[\hat{p}]_{(-s)}}(\hat{t}^{(s)}_{nm})\nonumber\\
&=& \frac{1}{4} \{ (1-s^{2})[\hat{q}\hat{p}\hat{t}^{(s)}_{nm}
+\hat{t}^{(s)}_{nm}\hat{p}\hat{q}]+
(1+s)^{2}\hat{q}\hat{t}^{(s)}_{nm}\hat{p}+
(1-s)^{2}\hat{p}\hat{t}^{(s)}_{nm}\hat{q} \}.
\end{eqnarray}

Finally in this section, with a phase space function expandable as power
series in $p$ and $q$ 
\begin{eqnarray}
f(q,p)=\sum_{n,m}c_{nm}q^{n}p^{m},
\end{eqnarray}
we associate a Liouvillian superoperator 
\begin{eqnarray}
\hat{\bf f}^{(s)}=\hat{\bf f}(\frac{1}{2}\hat{\bf T}_{[\hat{q}]_{(s)}},
\frac{1}{2}\hat{\bf T}_{[\hat{p}]_{(-s)}})=
\sum_{n,m}c_{nm}\hat{\bf O}^{(s)}_{nm}.
\end{eqnarray}
Note that like $f(q,p)$, $\hat{{\bf f}}^{(s)}$ is also a commutative
function of its arguments, and in this sense the Liouvillian superoperators
defined here mimic the fundamental property of the corresponding
phase space functions in
${\cal {F}(H)}$. $\hat{{\bf O}}^{(s)}_{nm}$ is the Liouvillian superoperator
corresponding to the monomial $q^{n}p^{m}$. For short, we write $\hat{{\bf f}%
}^{(s)}, \hat{{\bf O}}^{(s)}_{nm}$ without denoting their arguments.

\section{WWGM Quantization}

Let us denote by $N=C^\infty (M)$ the vector space of functions defined over
a phase space $M$ and by ${\cal {F}(H)}$ the vector space of operators
acting in a Hilbert space ${\cal {H}}$. While with the usual pointwise
product $N$ becomes a commutative and associative algebra, ${\cal {F}(H)}$
becomes a noncommutative but associative algebra with respect to usual
operator product. A noncommutative but associative algebra structure on $N$
can be implemented by $\star -$product; $\star _{(-s)}:N\times N\rightarrow N
$, explicitly given by \cite{Note1} 
\begin{eqnarray}
\star_{(-s)}=\exp \frac {1}{2}i\hbar
[(1-s)\partial^{L}_{p}\partial^{R}_{q}-
(1+s)\partial^{L}_{q}\partial^{R}_{p}].
\end{eqnarray}
Here we take $(q, p) \in {\bf R}^{2}=M$ and use the convention that $\partial ^L$ and $\partial ^R$ are acting
on the left (L) and on the right (R), respectively. Thus two different Lie
algebras structure can be defined on $N$; with respect to PB; $%
\{,\}_{PB}:N\times N\rightarrow N$ defined by 
\begin{eqnarray}
\{f,g\}_{PB}=\partial_{p}f\partial_{q}g-\partial_{q}f\partial_{p}g,
\end{eqnarray}
(henceforth the notation $\partial _x\equiv \partial /\partial x$ will be
used) and with respect to $s-$MB defined by 
\begin{eqnarray}
\{f_{1}(q,p),f_{2}(q,p)\}^{(-s)}_{MB}\equiv
f_{1}(q,p)\star_{(-s)}f_{2}(q,p)-f_{2}(q,p)\star_{(-s)}f_{1}(q,p).
\end{eqnarray}
where $f_1,f_2\in N$. Let us denote these two Lie algebras by $N_{PB}$ and $%
N_{MB}$, where the subscribes PB and MB refer to the respective brackets.
The relations (16) and (18) give the different expressions for the star
product and Moyal brackets, that appeared in the literature separately in a
unified manner and generalize them for an arbitrary s-ordering \cite{Vercin}.

Despite these two different Lie algebra structure there is only one in $%
{\cal {F}(H)}$ defined with respect to the usual Lie Bracket $[,]$, which we
denote by ${\cal {F}}_{LB}$. This is (anti-)homomorphic to Lie algebra
$N_{MB}$:
$\{q^{n}p^{m}, q^{k}p^{l}\}^{(-s)}_{MB} \rightarrow -[\hat{t}_{nm}^{(s)}, \hat{t}_{kl}^{(s)}]$
(see Eq. (61) of the second paper of Ref. \cite{Vercin}).
In the next section we will obtain a new quantum bracket which,
quite in parallel with the Lie algebra structures in $N$, enables us to
define a new Lie algebra structure in ${\cal {F}(H)}$. Before doing that we
have to recall some fundamental relations of the WWGM-quantization.

The above mentioned (anti-)homomorphism between $N_{MB}$ and ${\cal {F}}_{LB}$
is established via WWGM quantization rule symbolically defined by linear and
invertible map ${\cal M}_{s}:N\rightarrow {\cal {F}(H)}$ with inverse ${\cal %
M}_s^{-1}:{\cal {F}(H)} \rightarrow N$, such that ${\cal M}_{s}{\cal M}%
_{s}^{-1}$ and ${\cal M}_s^{-1}{\cal M}_{s}$ are identity transformations on 
${\cal {F}(H)}$ and $N$, respectively. Explicitly we write ${\cal M}_{s}(f)=%
\hat{F}^{(s)}$, and ${\cal M}_s^{-1}(\hat{F}^{(s)})=f$ where \cite{Note2} 
\begin{eqnarray}
\hat{F}^{(s)}(\hat{q},\hat{p})=
h^{-1} \int\int f(q,p)\hat{\Delta}_{qp}(s)dqdp;\qquad
f(q,p)=Tr[\hat{F}^{(s)}\hat{\Delta}_{qp}(-s)]
\end{eqnarray}
(All the integrals are from $-\infty $ to $\infty $). The first relation is
an expansion of an operator in a complete continuous operator basis 
\begin{eqnarray}
\hat{\Delta}_{qp}(s)=(\hbar/2\pi)\int\int
e^{-i(\xi q+\eta p)}\hat{D}(s)d\xi d\eta,
\end{eqnarray}
obeying the relations 
\begin{eqnarray}
\int \int \hat{\Delta}_{qp}(s) dqdp=h,\qquad Tr[\hat{\Delta}_{qp}(s)]=1.
\end{eqnarray}
Here $\hat D(s)=e^{-i\hbar s\xi \eta /2}\exp i(\xi \hat q+\eta \hat p)$ is
the $s-$parametrized displacement operator. The basis operators $\hat{\Delta}%
_{qp}$ are known as the Grossmann-Royer displaced parity operators \cite
{Grossmann} for $s=0$ and as the Kirkwood bases for $s=\pm 1$. Since they
form complete operator bases, in the sense that any operator obeying certain
conditions can be expanded in terms of them as in the first relation given
by (19), they provide a unified approach to different quantization rules 
\cite{Cahill,Balazs}. For special values $s=1,0,-1$ these are known,
respectively, as the standard, the Wigner-Weyl, and the antistandard rules
of associations \cite{Hillery,Balazs}.

The second relation in Eq. (19) easily follows by multiplying both sides of
the first relation by $\hat{\Delta}_{q^{\prime} p^{\prime}}(-s)$, and making
use of the relation 
\begin{eqnarray}
Tr[\hat{\Delta}_{qp}(s)\hat{\Delta}_{q^{\prime} p^{\prime}}(-s)] =
h\delta(q-q^{\prime})\delta(p-p^{\prime}).
\end{eqnarray}
Among other nice properties of the $\hat{\Delta}(s)$ basis we quote the so
called differential properties 
\begin{eqnarray}
\partial_{q} \hat{\Delta}_{qp}(s)&=&
-\frac{i}{\hbar} [\hat{p}, \hat{\Delta}_{qp}(s)]
,\qquad \partial_{p} \hat{\Delta}_{qp}(s) =
\frac{i}{\hbar} [\hat{q}, \hat{\Delta}_{qp}(s)]\\
q\hat{\Delta}_{qp}(s)&=&
\frac{1}{2}{\bf \hat{T}_{[\hat{q}]_{(s)}}}\hat{\Delta}_{qp}(s)
,\qquad p\hat{\Delta}_{qp}(s) =
\frac{1}{2}{\bf \hat{T}_{[\hat{p}]_{(-s)}}}\hat{\Delta}_{qp}(s).
\end{eqnarray}
These last relations can be generalized as 
\begin{eqnarray}
q^{n}p^{m}\hat{\Delta}_{qp}(s)=\hat{\bf O}^{(s)}_{nm}(\hat{\Delta}_{qp}(s)).
\end{eqnarray}
As an illustration, taking the traces of both sides we have $q^{n}p^{m}=Tr[%
\hat{t}_{nm}^{(s)}\hat{\Delta}_{qp}(-s)]$ which shows that ${\cal M}%
_{s}(q^{n}p^{m})=\hat{t}_{nm}^{(s)}$, or ${\cal M}^{-1}_{s}(\hat{t}%
_{nm}^{(s)})=q^{n}p^{m}$. More generally, for a function accepting power
series expansion as in (14) we see that the corresponding operator in $s-$%
association given by (19) is obtained by simply replacing $q^{n}p^{m}$ by $%
\hat{t}_{nm}^{(s)}$. For these kind of functions a generalization of (25) is 
\begin{eqnarray}
f(q, p)\hat{\Delta}_{qp}(s)=\hat{\bf f}^{(s)}(\hat{\Delta}_{qp}(s)).
\end{eqnarray}

Now by multiplying both sides of this relation by another function $g(q,p)$
we have 
\begin{eqnarray}
g(q, p)f(q, p)\hat{\Delta}_{qp}(s)=
\hat{\bf f}^{(s)}[\hat{\bf g}^{(s)}(\hat{\Delta}_{qp}(s))]
=\hat{\bf g}^{(s)}[\hat{\bf f}^{(s)}(\hat{\Delta}_{qp}(s))].\nonumber
\end{eqnarray}
By taking the integral and trace of all sides and making use of the
relations (21) we arrive at 
\begin{eqnarray}
h^{-1}\int \int g(q, p)f(q, p)\hat{\Delta}_{qp}(s)dqdp=
\hat{\bf f}^{(s)}[\hat{\bf g}^{(s)}(\hat{I})]
=\hat{\bf g}^{(s)}[\hat{\bf f}^{(s)}(\hat{I})]
=\hat{\bf f}^{(s)}(\hat{G}^{(s)})=\hat{\bf g}^{(s)}(\hat{F}^{(s)}),\\
g(q, p)f(q, p)=Tr\{[\hat{\bf g}^{(s)}(\hat{\bf F}^{(s)})]\hat{\Delta}_{qp}(-s)\}
=Tr\{[\hat{\bf f}^{(s)}(\hat{G}^{(s)})]\hat{\Delta}_{qp}(-s)\}.
\end{eqnarray}
These two relations explicitly answer the first question stated in the
introduction. Under the WWGM-association corresponding to an arbitrary
s-ordering, to the product of two c-number functions there corresponds an
operator which results by action of the Liouvillian superoperator form of
one on the other. More formally, in accordance with (19) we obtain 
\begin{eqnarray}
{\cal M}_{s}[g(q, p)f(q, p)]=\hat{\bf g}^{(s)}(\hat{F}^{(s)})
=\hat{\bf f}^{(s)}(\hat{G}^{(s)}).
\end{eqnarray}
As an example, by making use of Eq. (11) we have 
\begin{eqnarray*}
{\cal M}_{s}(q^{n_1}p^{m_1}q^{n_2}p^{m_2})=
\hat {{\bf O}}_{n_1m_1}^{(s)}(\hat t_{n_2m_2}^{(s)})=
\hat {{\bf O}}_{n_2m_2}^{(s)}(\hat t_{n_1m_1}^{(s)})=
\hat t_{n_1+n_2,m_1+m_2}^{(s)}.
\end{eqnarray*}
Note that the result
is, in general, different from the noncommutative product
$\hat t_{n_2m_2}^{(s)}\hat t_{n_1m_1}^{(s)}$, or from
$\hat t_{n_1m_1}^{(s)}\hat t_{n_2m_2}^{(s)}$.

\section{Derivation of the new Bracket}

Taking the derivatives of the second relation in Eq. (19) with respect to q
(and p) and then making use of Eq. (23) we have 
\begin{eqnarray}
\partial_{p}f(q,p)&=&Tr[\hat{F}^{(s)}\partial_{p}(\hat{\Delta}_{qp}(-s))]\nonumber\\
&=& -\frac{i}{\hbar}Tr[({\bf ad_{\hat{q}}}\hat{F}^{(s)})\hat{\Delta}_{qp}(-s)],\\
\partial_{q}g(q,p)&=&Tr[\hat{G}^{(s)}\partial_{q}(\hat{\Delta}_{qp}(-s))]\nonumber\\
&=& \frac{i}{\hbar}Tr[({\bf ad_{\hat{p}}}\hat{G}^{(s)})\hat{\Delta}_{qp}(-s),
\end{eqnarray}
where ${\bf ad_{\hat A}}$ denotes the adjoint action: ${\bf ad_{\hat A}}\hat 
B=[\hat A,\hat B]$. These relations show that, if ${\cal M}_{s}(f)=\hat{F}%
^{(s)} $, then ${\cal M}_{s}(\partial _pf)=-(i/\hbar ){\bf ad_{\hat q}}\hat{F%
}^{(s)} $ and ${\cal M}_{s}(\partial _qf)=(i/\hbar ){\bf ad_{\hat p}}\hat{F}%
^{(s)}$. Now by multiplying both sides of Eq. (30) by $\partial _qg$ and
making use of (26) and then of (12) we have 
\begin{eqnarray}
\partial_{p}f\partial_{q}g&=&
-\frac{i}{\hbar}Tr[({\bf ad_{\hat{q}}}\hat{F}^{(s)})
{\bf \hat{g}_{q}^{(-s)}}(\hat{\Delta}_{qp}(-s))],\nonumber\\
&=&-\frac{i}{\hbar}
Tr[\hat{\Delta}_{qp}(-s){\bf \hat{g}_{q}^{(s)}}({\bf ad_{\hat{q}}}\hat{F}^{(s)})].
\end{eqnarray}
By reversing the order of manipulations these relations can be rewritten as 
\begin{eqnarray}
\partial_{p}f\partial_{q}g=\frac{i}{\hbar}
Tr[\hat{\Delta}_{qp}(-s){\bf \hat{f}_{p}^{(s)}}({\bf ad_{\hat{p}}}\hat{G}^{(s)})],
\end{eqnarray}
where ${\bf \hat h_x^{(s)}}$ stands for the superoperator associated to the $%
\partial _xh$. In a similar way we have 
\begin{eqnarray}
\partial_{q}f\partial_{p}g&=&\frac{i}{\hbar}
Tr[{\bf \hat{g}_{p}^{(s)}}({\bf ad_{\hat{p}}}\hat{F}^{(s)})\hat{\Delta}_{qp}(-s)]\\
&=&-\frac{i}{\hbar}
Tr[{\bf \hat{f}_{q}^{(s)}}({\bf ad_{\hat{q}}}\hat{G}^{(s)})\hat{\Delta}_{qp}(-s)].
\end{eqnarray}
Thus by combining Eqs. (32, 33) with (34, 35) we arrive at four differently
looking but equivalent expressions for PB of two functions: 
\begin{eqnarray}
\{f,g\}_{PB}&=&-\frac{i}{\hbar}
Tr\{[{\bf \hat{g}_{q}^{(s)}}({\bf ad_{\hat{q}}}\hat{F}^{(s)})
+{\bf \hat{g}_{p}^{(s)}}({\bf ad_{\hat{p}}}\hat{F}^{(s)})]\hat{\Delta}_{qp}(-s)\}\\
&=&-\frac{i}{\hbar}Tr\{[{\bf \hat{g}_{q}^{(s)}}({\bf ad_{\hat{q}}}\hat{F}^{(s)})
-{\bf \hat{f}_{q}^{(s)}}({\bf ad_{\hat{q}}}\hat{G}^{(s)})]\hat{\Delta}_{qp}(-s)\}\\
&=&\frac{i}{\hbar}Tr\{[{\bf \hat{f}_{p}^{(s)}}({\bf ad_{\hat{p}}}\hat{G}^{(s)})
-{\bf \hat{g}_{p}^{(s)}}({\bf ad_{\hat{p}}}\hat{F}^{(s)})]\hat{\Delta}_{qp}(-s)\}\\
&=&\frac{i}{\hbar}Tr\{[{\bf \hat{f}_{q}^{(s)}}({\bf ad_{\hat{q}}}\hat{G}^{(s)})
+{\bf \hat{f}_{p}^{(s)}}({\bf ad_{\hat{p}}}\hat{G}^{(s)})]\hat{\Delta}_{qp}(-s)\}.
\end{eqnarray}
These relations enable us to define a new bracket in ${\cal {F}(H)}$ which
we denote by $[,]^{(s)}_{PMB}$, and call it the Poisson-Moyal bracket (PMB).
It is defined as the image of the PB under the WWGM-association: 
\begin{eqnarray}
{\cal M}_{s}(\{f,g\}_{PB})=[\hat{F}, \hat{G}]^{(s)}_{PMB},
\end{eqnarray}
and explicitly given by the following four equivalent expressions 
\begin{eqnarray}
[\hat{F}, \hat{G}]^{(s)}_{PMB}
&=&-\frac{i}{\hbar}[{\bf \hat{g}_{q}^{(s)}}({\bf ad_{\hat{q}}}\hat{F}^{(s)})
+{\bf \hat{g}_{p}^{(s)}}({\bf ad_{\hat{p}}}\hat{F}^{(s)})]\\
&=&-\frac{i}{\hbar}[{\bf \hat{g}_{q}^{(s)}}({\bf ad_{\hat{q}}}\hat{F}^{(s)})
-{\bf \hat{f}_{q}^{(s)}}({\bf ad_{\hat{q}}}\hat{G}^{(s)})]\\
&=&\frac{i}{\hbar}[{\bf \hat{f}_{p}^{(s)}}({\bf ad_{\hat{p}}}\hat{G}^{(s)})
-{\bf \hat{g}_{p}^{(s)}}({\bf ad_{\hat{p}}}\hat{F}^{(s)})]\\
&=&\frac{i}{\hbar}[{\bf \hat{f}_{q}^{(s)}}({\bf ad_{\hat{q}}}\hat{G}^{(s)})
+{\bf \hat{f}_{p}^{(s)}}({\bf ad_{\hat{p}}}\hat{G}^{(s)})].
\end{eqnarray}
Obviously, since the WWGM-association is linear, and PB is a Lie bracket
i.e., bilinear, antisymmetric and obeying Jacobi identity, so is this new
PMB. The four seemingly different but equivalent expressions of this new
bracket correspond to trivially equivalent rearrangement of the terms in the
right hand side of (17). In contrast, the equivalences of the Eqs. (41-44) are
not so trivial.

In the case of many degrees of freedom, in the right hand sides of Eqs.
(41-44) $\hat q$ and $\hat p$ are to be labeled with an index and summed
over them. For instance, the second and third ones are to be as follows 
\begin{eqnarray}
[\hat{F}, \hat{G}]^{(s)}_{PMB}
&=&-\frac{i}{\hbar} \sum _{i}
[{\bf \hat{g}_{q_{i}}^{(s)}}({\bf ad_{\hat{q_{i}}}}\hat{F}^{(s)})
-{\bf \hat{f}_{q_{i}}^{(s)}}({\bf ad_{\hat{q_{i}}}}\hat{G}^{(s)})]\\
&=&\frac{i}{\hbar} \sum _{i}
[{\bf \hat{f}_{p_{i}}^{(s)}}({\bf ad_{\hat{p_{i}}}}\hat{G}^{(s)})
-{\bf \hat{g}_{p_{i}}^{(s)}}({\bf ad_{\hat{p_{i}}}}\hat{F}^{(s)})].
\end{eqnarray}

\section{APPLICATIONS}

As a general first application we take $f(q,p)=q^{n}p^{m}$ and $%
g(q,p)=q^{k}p^{l}$; $n,m,k,l$ integers. These kind of monomials form a basis
for the so called $w_{\infty}$-algebra with respect to PB: 
\begin{eqnarray}
\{q^{n}p^{m}, q^{k}p^{l}\}_{PB}=(mk-nl)q^{n+k-1}p^{m+l-1},
\end{eqnarray}
and $W_{\infty}$ algebra with respect to $s-MB$ \cite{Vercin} 
\begin{eqnarray}
\{q^{n}p^{m},q^{k}p^{l}\}^{(-s)}_{MB}=
\sum^{j_{max}}_{j=0}\frac{i^{j}}{j!}[\sum^{j \prime}_{r=0}(^{j}_{r})
f_{srj}a_{nmkl,rj}]q^{n+k-j}p^{m+l-j}.
\end{eqnarray}
Here the prime over the second summation indicates that the maximum value
that $r$ may take is $r_{max}=(m,k)$ (i.e., the smaller of the integers $m$
and $k$) and 
\begin{eqnarray}
j_{max}=(n+r_{max},l+r_{max}) , \qquad
a_{nmkl,rj}=\frac{n!m!k!l!}{(n+r-j)!(m-r)!(k-r)!(l+r-j)!}.
\end{eqnarray}
The restrictions imposed on summations also follows from the expression of $%
a_{nmkl,rj}$. In Eq. (44) 
\begin{eqnarray}
f_{srj}=(s^{-})^{r}(-s^{+})^{j-r}-(s^{-})^{j-r}(-s^{+})^{r},\nonumber
\end{eqnarray}
is the only factor depending on the chosen rule of ordering, where $%
s^{\pm}=\hbar (1 \pm s)/2$. The $w_{\infty}-$algebra is the algebra of
canonical diffeomorphisms of a phase space that is topologically equivalent
to ${\bf {R}^{2}}$, or, since the area element and symplectic form coincide
in two dimensions, as the algebra of area preserving diffeomorphisms $%
Diff_{A}{\bf {R}^{2}}$ \cite{Pope}. The above given $W_{\infty}$-algebra is
quantum (or, $\hbar$) deformation of this classical $w_{\infty}$. More
explicitly, one can easily show that 
\begin{eqnarray}
\lim _{\hbar \rightarrow 0}  (i\hbar)^{-1}\{ , \}^{(-s)}_{MB}=\{ , \}_{PB}.
\end{eqnarray}

Now with respect to our new bracket we will obtain an algebra isomorphic to $%
w_{\infty}$-algebra. This will be denoted by ${\cal {F}}_{PMB}$. From (19)
we obtain 
\begin{eqnarray}
\hat{F}^{(s)}=\hat{t}_{nm}^{(s)}, \qquad \hat{G}^{(s)}=\hat{t}_{kl}^{(s)},
\end{eqnarray}
and by making use of (7) 
\begin{eqnarray}
{\bf ad_{\hat{q}}}\hat{F}^{(s)}&=&i\hbar m\hat{t}_{n,m-1}^{(s)}, \qquad
{\bf ad_{\hat{p}}}\hat{F}^{(s)}=-i\hbar n\hat{t}_{n-1,m}^{(s)},\\
{\bf ad_{\hat{q}}}\hat{G}^{(s)}&=&i\hbar l\hat{t}_{k,l-1}^{(s)}, \qquad
{\bf ad_{\hat{p}}}\hat{G}^{(s)}=-i\hbar k\hat{t}_{k-1,l}^{(s)}.
\end{eqnarray}
The corresponding superoperators are as follows 
\begin{eqnarray}
\hat{\bf f}^{(s)}_{q}&=& n 2^{-(n+m-1)}\hat{\bf T}^{n-1}_{[\hat{q}]_{(s)}}
\hat{\bf T}^{m}_{[\hat{p}]_{(-s)}}=n \hat{\bf O}^{(s)}_{n-1, m}, \nonumber\\
\hat{\bf f}^{(s)}_{p}&=& m\hat{\bf O}^{(s)}_{n, m-1},\nonumber\\
\hat{\bf g}^{(s)}_{q}&=& k\hat{\bf O}^{(s)}_{k-1, l}, \qquad
\hat{\bf g}^{(s)}_{p} = l \hat{\bf O}^{(s)}_{k, l-1}.
\end{eqnarray}
Substituting these relations in anyone of that given in Eq. (41-44) and by
using the identities such as (see Eq. (11)) 
\begin{eqnarray}
\hat{\bf O}^{(s)}_{k-1, l}(\hat{t}^{(s)}_{n, m-1})=
\hat{\bf O}^{(s)}_{k, l-1}(\hat{t}^{(s)}_{n-1, m})=
\hat{t}^{(s)}_{n+k-1, m+l-1},\nonumber
\end{eqnarray}
we obtain 
\begin{eqnarray}
[\hat{F}, \hat{G}]^{(s)}_{PMB}=(mk-nl)\hat{t}_{n+k-1, m+l-1}^{(s)}.
\end{eqnarray}
Thus, by comparing with (47) we see that ${\cal F}_{PMB}$ is
isomorphic to $w_{\infty}$-algebra. Because of (50), or, as can be directly
verified, we have 
\begin{eqnarray}
-\lim _{\hbar \rightarrow 0}  (i\hbar)^{-1}[ , ]=[ , ]_{PMB}.
\end{eqnarray}
provided that the same ordering convention is used in both sides.

There are some remarkable particular cases of this general application that
deserve to be mentioned. $W_{\infty}$-algebra has some abelian and finite
or infinite dimensional nonabelian subalgebras for which structure constants
are proportional to the first power of $i \hbar$ \cite{Vercin}. These are
generated by $\hat{t}_{nm}^{(s)}$ such that : (i)$n=0$, (ii)$m=0$ , (iii)$%
n=m $ (Cartan subalgebra)\cite{Vercin1}, (iv)$n+m \leq 1$ (HW-algebra), (v)$%
n+m=2 $ (symplectic algebra $sp(2)$), (vi) $n+m \leq 2$ (inhomogeneous
symplectic algebra $isp(2)$), (vii) $m=1$, (viii)$n=1$. The first three are
infinite dimensional abelian subalgebras and the last two are isomorphic
copy of the well known centerless Virasoro algebra \cite{Pope}. For all
these subalgebras Eq. (56) is of the form $-(i\hbar)^{-1}[ , ]=[ , ]_{PMB}$.

As a second general application we will carry the classical Hamiltonian
equations of motion 
\begin{eqnarray}
\dot{q}=-\{q, H\}_{PB}=\partial_{p}H,\qquad \dot{p}=
-\{p, H\}_{PB}=-\partial_{q}H
\end{eqnarray}
to ${\cal {F}(H)}$. Here $H\equiv H(q, p)$ is the classical Hamiltonian and $%
t$ being the time parameter $\dot{a}\equiv da/dt$. Now applying ${\cal {M}}%
_{s}$ to both sides of Eq. (57) we obtain 
\begin{eqnarray}
\dot{\hat{q}}=\frac{1}{i \hbar}[\hat{q}, \hat{H}^{(s)}],\qquad
\dot{\hat{p}}=\frac{1}{i \hbar}[\hat{p}, \hat{H}^{(s)}]
\end{eqnarray}
here $\hat{H}^{(s)}={\cal {M}}_{s}(H)$, and we used the fact that 
\begin{eqnarray}
[\hat{q},\hat{H}]^{(s)}_{PMB}=(i/\hbar)[\hat{q}, \hat{H}^{(s)}], \qquad
[\hat{q},\hat{H}]^{(s)}_{PMB}=(i/\hbar)[\hat{p}, \hat{H}^{(s)}].
\end{eqnarray}
Notice that $\hat{H}^{(s)}$ is Hermitian only for pure imaginary values of $s
$, that is $(\hat{H}^{(s)})^{\dagger}=\hat{H}^{(-\bar{s})}$. In particular,
when $H=(p^{2}/2m)+V(q)$ Eqs. (58) are of the form (Ehrenfest's Theorem) 
\begin{eqnarray}
\dot{\hat{q}}=\frac{\hat{p}}{m},\qquad
\dot{\hat{p}}= -\frac{i}{\hbar}[\hat{p}, \hat{V}(\hat{q})].
\end{eqnarray}
Note that these are independent from $s$.

We would like to call Eqs. (58) the operator form of the Hamilton equations.
Assume that the operators belong to Heisenberg picture, these equations are
identical to Heisenberg picture equations of motion that can be obtained
from 
\begin{eqnarray}
\frac{d \hat{A}_{H}}{dt}=\frac{\partial \hat{A}_{H}}{\partial t}
+ \frac{1}{i \hbar}[\hat{A}_{H}, \hat{H}],
\end{eqnarray}
by taking $\hat A_H=\hat q,\hat p$ and $\hat{H}=\hat{H}^{(s)}$. Here the
subscribe $H$ refers to the Heisenberg picture in which the state vectors
are time-independent and the dynamical variables are time-dependent. We
should note that the first term in the right side of Eq. (61) is defined as
follows \cite{Schiff} 
\begin{eqnarray}
\frac{\partial \hat{A}_{H}}{\partial t}\equiv
(\frac{\partial \hat{A}}{\partial t})_{H}
=\hat{U}\frac{\partial \hat{A}_{S}}{\partial t}
\hat{U}^{\dagger},
\end{eqnarray}
where $\hat U=\exp (it\hat H/\hbar )$ is the evolution operator and the
subscribe $S$ refers to the Schr\"odinger picture in which the state vectors
are time-dependent and dynamical variables are time-independent (except for
a possible explicit time dependence, which is not the case for the position
and momentum operators in the Schr\"odinger picture).

As a result, as far as the dynamics of $\hat{q}$ and $\hat{p}$ are
concerned, the WWGM-association directly maps the classical Hamilton
equations of motion on to the Heisenberg picture equations of motion for
general Hamiltonian $\hat{H}^{(s)}={\cal M}_{s}(H)$. Arrival to the
Schr\"odinger equation (see the next section) for the time evolution of
states vectors is straightforward by making use of the evolution operator $%
\hat U$ in the case of $\bar{s}=-s$. Despite of this application, the fact
that the image of the classical mechanics under the WWGM-association is not
identical to the Heisenberg picture formulation of the conventional quantum
mechanics is made apparent in the next application.

Finally, we consider the equation describing the time evolution of a phase
space function $f\equiv f(q,p;t)$ 
\begin{eqnarray}
\dot{f}=\partial_{t}f +\{H,f\}_{PB},
\end{eqnarray}
associated with a system described by $H$. The corresponding equation in $%
{\cal {F}(H)}$ is as follows 
\begin{eqnarray}
\dot{\hat{F}}^{(s)}=\partial_{t} \hat{F}^{(s)} + [\hat{H}, \hat{F}]^{(s)}_{PMB},
\end{eqnarray}
where $\hat{F}^{(s)}={\cal {M}}_{s}(f)$. In particular, for $H=(p^2/2m)+V(q)$
, $f=f(q)$ and $g=g(p)$ the equations are 
\begin{eqnarray}
\dot{\hat{F}}^{(s)}=\frac{1}{m}{\bf \hat{f}^{(s)}_{q}}(\hat{p}),\qquad
\dot{\hat{G}}^{(s)}=-\frac{i}{\hbar}{\bf \hat{g}^{(s)}_{p}}[\hat{p},\hat{V}(\hat{q})].
\end{eqnarray}
Note that the operator form of Hamilton equations are particular case of
these last equations. The distinction between the Heisenberg picture
formulation of the quantum mechanics and the image of the Hamilton
formulation of the classical mechanics is made manifest by (64) by
appearence of PMB instead of $[,]$.

\section{conclusion and discussion}

The conventional way for finding a quantum system that reduce to a specified
classical system in the classical limit is to write the classical Hamilton
equations in terms of the PB and then to replace the PB with commutator
brackets in accordance with $\{f,g\}_{PB}\rightarrow (i\hbar )^{-1}[\hat F,%
\hat G]$. Although this construction suffers from the obstructions stated in
the introduction, the association $\{f,g\}_{PB}\rightarrow [\hat F,\hat G%
]^{(s)}_{PMB}$ is free from them. Note that $-(i \hbar)^{-1}[,]^{(s)}_{PMB}$
reduces to the commutator bracket when one of the entry is $\hat q$, or $%
\hat p$. A bit more generally, when $\hat F=a\hat q+b\hat p+c\hat I$, a
general element of the HW-algebra, then $-[\hat F,\hat H]^{(s)}_{PMB}=(i%
\hbar )^{-1}[\hat F,\hat{H}^{(s)}]$. Thus, for time-independent Hamiltonian,
as the solutions of Eqs. (58), the time evolution of the basic observables
are given by $\hat q(t)=\hat U(t,s)\hat q(0)\hat U(-t,s),  \hat p(t)=\hat U%
(t,s)\hat p(0)\hat U(-t,s)$. Here $\hat q(0)$, and $\hat p(0)$ are time
independent position and momentum operators and $\hat{U}(t,s)=\exp (i t 
\hat{H}^{(s)}/ \hbar)$. By noting that $\hat{U}(t,s)$ is unitary only when $%
\bar{s}=-s$, these are the same as that in the Heisenberg picture if $\hat{q}%
(0)$ and $\hat{p}(0)$ are considered to be in the Schr\"odinger picture and
if $\bar{s}=-s$. In that case, by assuming time-independent state vector $%
|\psi (0)>\in {\cal H}$ in the Heisenberg picture such that $<\psi (0)|\hat{p%
}(t)|\psi(0)>=<\psi(t)|\hat{p}(0)|\psi (t)>$, we obtain time-dependent state
vector $|\psi (t)>=\hat{U}(-t,s)|\psi (0)>$ obeying the dynamics $i \hbar
\partial_{t}|\psi (t)>=\hat{H}^{(s)}|\psi (t)>$, i.e., the Schr\"odinger
equation.

On the other hand, while the time evolution of a general observable (not
explicitly time-dependent) is given in the Heisenberg picture by $\hat{F}%
_{H}(t)=\hat U(t)\hat{F}_{H}(0)\hat U(t)^{\dagger }$, it is not so in the
association scheme $PB \rightarrow PMB$. Instead, if we define
$({\bf ad_{\hat A}})^{(s)}_{PM}$ by
$({\bf ad_{\hat A}})^{(s)}_{PM}\hat B=[\hat A,\hat B]^{(s)}_{PMB}$, then
the time evolution governed by Eq. (64) can be written as
$\hat{F}^{(s)}(t)=[\exp {-t({\bf ad_{\hat H}})^{(s)}_{PM}}]\hat{F}^{(s)}(0)$
which, because of (50) or (56), is the limiting case of the above
given relation.

In the sense described above, the conventional canonical quantization itself
can be thought as an $\hbar $ deformation of the quantization by PMB. This
fact is made manifest by the following diagrams:

(i)\underline{Hiearachy of Products}

\begin{eqnarray*}
f.g=g.f & \Leftarrow WWGM-association \Rightarrow &
\hat{F}\Diamond \hat{G}=\hat{G}\Diamond \hat{F}\\
def.\downarrow \uparrow cont. &  & def.\downarrow \uparrow cont.\\
f\star g \neq g\star f & \Leftarrow WWGM-association \Rightarrow &
\hat{F}\hat{G} \neq \hat{G}\hat{F} 
\end{eqnarray*}

(ii)\underline{Hiearachy of Brackets}

\begin{eqnarray*}
\{., .\}_{PB} \Leftarrow & WWGM-association & \Rightarrow  [., .]_{PMB}\\
def.\downarrow \uparrow cont. &  & def.\downarrow \uparrow cont.\\
\{., .\}_{MB} \Leftarrow & WWGM-association & \Rightarrow  [., .]\\ 
\end{eqnarray*}
These two diagrams schematically summarize the main points of this report,
and exhibit the hierarchies of the products and brackets involved. Here
cont. and def. stand for contraction and deformation, respectively and, for
the sake of simplicity the commutative operator product derived in Eq. (29)
is shown here by $\hat {{\bf f}}(\hat G)\equiv \hat F\Diamond \hat G$. Note
that the $\Diamond $ product (like the $\star $ product and the MB) depends
on the ordering parameter $s$, but this is not shown in the diagrams for the
same reason. In summary, what we have done here may be considered as the
``contraction quantization'', or, as the formulation of the classical
mechanics in the ring of operators.

\acknowledgments
This report has been written during my visit to the Feza G\"ursey Institute
to which I would like to express my profound gratitude for the hospitality.
Also I wish to thank T. Dereli and \"O. F. Day\i\ for helpful discussions.
This work was supported in part by the Scientific and Technical Research
Council of Turkey (T\"UBITAK).

\end{document}